\documentclass[10pt]{article}
\usepackage[dvips]{graphics}
\usepackage{emulateapj}

\def\br{\hbox{$B\!-\!R$}}

\def\vi{\hbox{$V\!-\!I$}}
\newcommand{\hi}{H$\;${\small I}}
\newcommand{\kms}{\hbox{$\mathrm{km}\;\mathrm{s}^{-1}$}}

\begin{document}

\submitted{Accepted by The Astrophysical Journal Letters}

\title{Triple Bars and Complex Central Structures in Disk Galaxies}
\author{Peter Erwin and Linda S. Sparke}
\affil{University of Wisconsin-Madison}
\authoraddr{475 N. Charter St.\ \\
Madison WI 53706}

\begin{abstract}
We present an analysis of ground-based and HST images of three
early-type barred galaxies.  The first, NGC 2681, may be the clearest
example yet of a galaxy with three concentric bars.  The two other
galaxies were previously suggested as triple-barred.  Our analysis
shows that while NGC 3945 is probably double-barred, NGC 4371 has only
one bar; but both have intriguing central structures.  NGC 3945 has a
large, extremely bright disk inside its primary bar, with patchy dust
lanes, a faint nuclear ring or pseudo-ring within the disk, and an
apparent secondary bar crossing the ring.  NGC 4371 has a bright
nuclear ring only marginally bluer than the surrounding bulge and bar.
There is no evidence for significant dust or star formation in either
of these nuclear rings.  The presence of stellar nuclear rings
suggests that the centers of these galaxies are dynamically cool and
disklike.
\end{abstract}


\section{Introduction}
Many double-barred galaxies --- those with a smaller, concentric bar 
inside the main bar --- have been found in recent years; see 
\cite{friedli96b} for a review.  \cite{erwin99} find secondary bars in 
at least 20\% of their sample of barred S0--Sa field galaxies.  
Wozniak et al.\ (1995, hereafter \cite{w95}) and \cite{friedli96a} 
noted three galaxies that appeared to have \textit{three} concentric 
bar-like structures.  Bars-within-bars have been advocated as a way to 
bring gas into the nucleus of galaxies (e.g., \cite{shlosman}).  
\cite{pfenniger90} studied dissipative motion in a system with the 
inner bar rotating faster than the outer one, while 
\nocite{witold97}Maciejewski \& Sparke (1997) showed how these might 
be modeled as self-consistent dynamical systems, with particle orbits 
reproducing the shape of the potential.  Here, we present optical 
images from our survey with the 3.5m WIYN telescope and archival 
optical and near-IR images from the Hubble Space Telescope (HST) for 
two of \cite{w95}'s triple-bar candidates, NGC 3945 and NGC 4371, and 
a third galaxy, NGC 2681.\footnote{The WIYN Observatory is a joint 
facility of the University of Wisconsin-Madison, Indiana University, 
Yale University, and the National Optical Astronomy Observatories.  
Observations with the NASA/ESA Hubble Space Telescope obtained at the 
Space Telescope Science Institute, operated by the Association of 
Universities for Research in Astronomy, Inc., under NASA contract 
NAS5-26555.}

\section{Observations and Analysis}

Images were analyzed using the \textsc{iraf} ellipse package, which 
fits ellipses to isophotes (\cite{jz87}).  We identified bars by 
visual inspection of the images and by looking for peaks in isophote 
ellipticity that were accompanied by stationary values of the position 
angle, as described by \cite{w95}.  We also created unsharp masks, to 
look for subtle, small-scale brightness variations and structures not 
otherwise readily apparent.  Table~\ref{table1} lists the main 
features we found.

\textbf{NGC 2681, Figure~1:} This galaxy is close to face-on; outer
disk isophotes have ellipticity $< 0.07$.  Thus, the three barlike
structures cannot be projection effects.  The outermost bar is
relatively large, though not outside the range for early-type galaxies
found by \cite{martin}.  Although this structure is relatively faint,
the PA is nearly stationary at $\approx 30\arcdeg$ for $a \approx
45$--$60\arcsec$ (1c).  At these radii, the isophotes have the
squared-off ends (1a, d) characteristic of bars in early-type galaxies
(\cite{athan90}).  Consequently, we feel reasonably confident that
this \textit{is} a bar.  The NICMOS image confirms the existence of a
third bar --- tentatively identified by \cite{w95} and
\cite{friedli96a} as the galaxy's secondary bar.

\textbf{NGC 3945, Figure~2:} Interpretation of structures in this 
galaxy depends strongly on what inclination is assumed.  \cite{w95} 
suggested that $i = 26\arcdeg$, and that the galaxy is triply barred.  
For two reasons, we feel that an inclination of $\sim 50\arcdeg$ is 
more likely, and that their secondary bar is probably an axisymmetric 
structure.  The primary bar ($a \approx 40\arcsec$) is surrounded by 
outer and inner rings (2d, g); both appear perpendicular to the bar.  
Inner rings are almost always elongated \textit{parallel} to the bar 
(\nocite{buta86}Buta 1986, 1995).  If the inner ring is intrinsically 
circular, then $i \approx 35\arcdeg$; $i \approx 50\arcdeg$ for a 
typical axis ratio (b/a = 0.78, \cite{buta95}).  The kinematics also 
imply a high inclination: \cite{kormendy82} measured stellar speeds of 
$\approx 250$ \kms in the lens region, while the \hi{} profile has a 
half-width of 340 \kms{} (\cite{wk86}).  If $i = 26\arcdeg$, then the 
circular velocity reaches an astonishing $V \sim 780$ \kms, higher 
even than in the fastest-rotating disk galaxy known, UGC 12591 
(\cite{giovanelli86}, $V \approx 500$ \kms).  For $i = 50\arcdeg$, we 
would have $V \sim 440$ \kms.

Our fits (2c) show \textit{seven} ellipticity peaks --- including the 
outer and inner rings --- all but two of which are within position 
angles 155--$160\arcdeg$.  At least some of the latter structures are 
probably near-axisymmetric, viewed at an inclination of $\sim 
50\arcdeg$.  \nocite{kormendy93}Kormendy (1993) has pointed out that 
the fast rotation and relatively low velocity dispersion in the 
``bulge'' region ($a < 20\arcsec$) of this galaxy suggest a disk 
rather than a bulge.  The structure \cite{w95} called a ``secondary 
bar'' (the ellipticity peak at $a \approx 10\arcsec$) is thus probably 
intrinsically round and flat --- an inner disk.  Assuming that it is 
vertically thin, the measured ellipticity of 0.36 deprojects to a 
circle if $i = 50\arcdeg$.

Unsharp masking (2e, f) shows a nuclear ring or tight spiral with $a 
\approx 6.5\arcsec$.  At $a \approx 2.5$--$3\arcsec$, the WFPC2 images 
show a possible foreshortened secondary bar crossing the ring.  The 
weak innermost ellipticity peak at $a < 2\arcsec$, \cite{w95}'s 
``tertiary bar'', shares the common system alignment; the galaxy 
center is probably mildly oblate rather than barred.  The \vi{} color 
map (2h) shows complex dust structures within the nuclear ring.

\textbf{NGC 4371, Figure~3:} \cite{w95} identified a ``secondary bar''
at $a \approx 10\arcsec$, aligned to within a few degrees of the outer
disk; they noted that it could be a projection effect, due to the
galaxy's high inclination.  Our unsharp mask (3e) shows that this
feature is a smooth nuclear ring, rather than a bar, with no signs of
dust or ongoing star formation.  \nocite{kormendy82}Kormendy (1982,
1993) noted disk-like dynamics in the bulge region; the ring may be
partially responsible.  The stellar ring is marginally bluer than the
surrounding bulge, or may be surrounded by a weak dust ring; cf.\ the
red ring noted by \cite{w95}.  The very center of the galaxy shows
twisted, spiral isophotes (3f), possibly due to a dusty nuclear disk.

\section{Discussion}

At HST resolution, NGC 2681 proves to have three distinct bars; NGC
3945 has two.  The relative orientation of the nested bars appears
random, as expected if the components rotate independently, and in
line with the findings of \cite{friedli96b}.  Both galaxies with
multiple bars have LINER nuclei, while NGC 4371 does not; this is weak
evidence in favor of \nocite{shlosman}Shlosman et al.'s (1989) thesis
that inner bars can enhance nuclear fuelling (but see
\cite{mulchaey}).

``Bars'' closely aligned with outer-disk isophotes must be treated
with suspicion: they may be flat, axisymmetric structures in
projection.  Combining unsharp masking with ellipse fits, we find a
nuclear ring in NGC 4371 and a bright disk-plus-ring in the inner
regions of NGC 3945.  Neither ring shows signs of current star
formation; they may be part of a previously overlooked population of
older, stellar nuclear rings in early-type galaxies.  A dynamically
``hot'' stellar system like a bulge, where the velocity dispersion
$\sigma$ is of the same magnitude as the rotation speed $V$, does not
form bars or rings; these are characteristic of ``cool'' disks, where
$V$ is at least a few times larger than $\sigma$.  So the wealth of
features revealed in these high-resolution images reinforces
\cite{kormendy93} suggestion: the bright inner regions of many
galaxies may be dominated by dynamically cool disks rather than
dynamically hot bulges.

\acknowledgements

We would like to thank Jay Gallagher for useful comments throughout 
this work, and the referee, Daniel Friedli, for helpful criticism.  
This research was supported by NSF grant AST-9320403, and NASA grants 
NAGW-2796 and AR-0798.01-96A from the Space Telescope Science 
Institute, operated by the Association of Universities for Research in 
Astronomy, Inc., under NASA contract NAS5-26555.



\newpage

\begin{deluxetable}{lllrrrrr}
\small
\footnotesize
\scriptsize
\tablecaption{\label{table1}}
\tablecolumns{8}
\tablewidth{0pt}
\tablehead{
\colhead{Galaxy} & \colhead{Type (RC3)} & \colhead{Nucleus} &
\multicolumn{5}{c}{Features} }
\startdata
NGC 2681 & (R$\arcmin$)SAB(rs)0/a & LINER & Outer Disk & Bar 1 & Bar 2 & Bar 3
& \nl
 & & & $r_{25} = 110\arcsec$ & $a = 50$--$60\arcsec$ & $15$--$20\arcsec$ &
$1.8$--$4.0\arcsec$ & \nl
 & & & PA = $130\arcdeg$ & $30\arcdeg$ & $75\arcdeg$ & $20\arcdeg$  & \nl
\tablevspace{5mm}
NGC 3945 & SB(rs)0$^{+}$ & LINER & Outer Ring & Bar 1 & Disk & Nuclear Ring & Bar 2 \nl
 & & & $a = 150\arcsec$ & $30$--$40\arcsec$ & $10\arcsec$ & $6.5\arcsec$ &
$2.5$--$3.0\arcsec$ \nl
 & & & PA = $160\arcdeg$ & $75\arcdeg$ & $158\arcdeg $ & $158\arcdeg$ &
$115\arcdeg$ \nl
\tablevspace{5mm}
NGC 4371 & SB(r)0$^{+}$ & No activity & Outer Disk  & Bar 1 & & Nuclear Ring & \nl
 & &  & $r_{25} = 120\arcsec$ & $a = 34$--$40\arcsec$ & & $10.5\arcsec$ & \nl
 & & & PA = $92\arcdeg$ & $158\arcdeg$ & & $92\arcdeg$ & \nl
\enddata
\tablecomments{Nuclear classifications are from \cite{ho97}, radii 
$r_{25}$ are from \cite{rc3}.  Position angle PA is measured from 
ellipse fits, where ellipticity is maximal ($\epsilon_{max}$).  
Semi-major axis $a$ is that at $\epsilon_{max}$ for disks and rings.  
Lower limit to bar length is at $\epsilon_{max}$; upper limit is the 
point outside outside $\epsilon_{max}$ where PA has changed by 
$10\arcdeg$.  Features with $a < 30\arcsec$ are measured using fits to 
HST images (see Figures 1c, 2c, 3c); larger features are measured using 
WIYN R-band images.  No deprojections were attempted.}
\end{deluxetable}

\newpage

\setlength{\oddsidemargin}{0cm}
\setlength{\evensidemargin}{0cm}
\setlength{\textwidth}{6.5in}

\begin{figure}
\scalebox{0.648}{\includegraphics{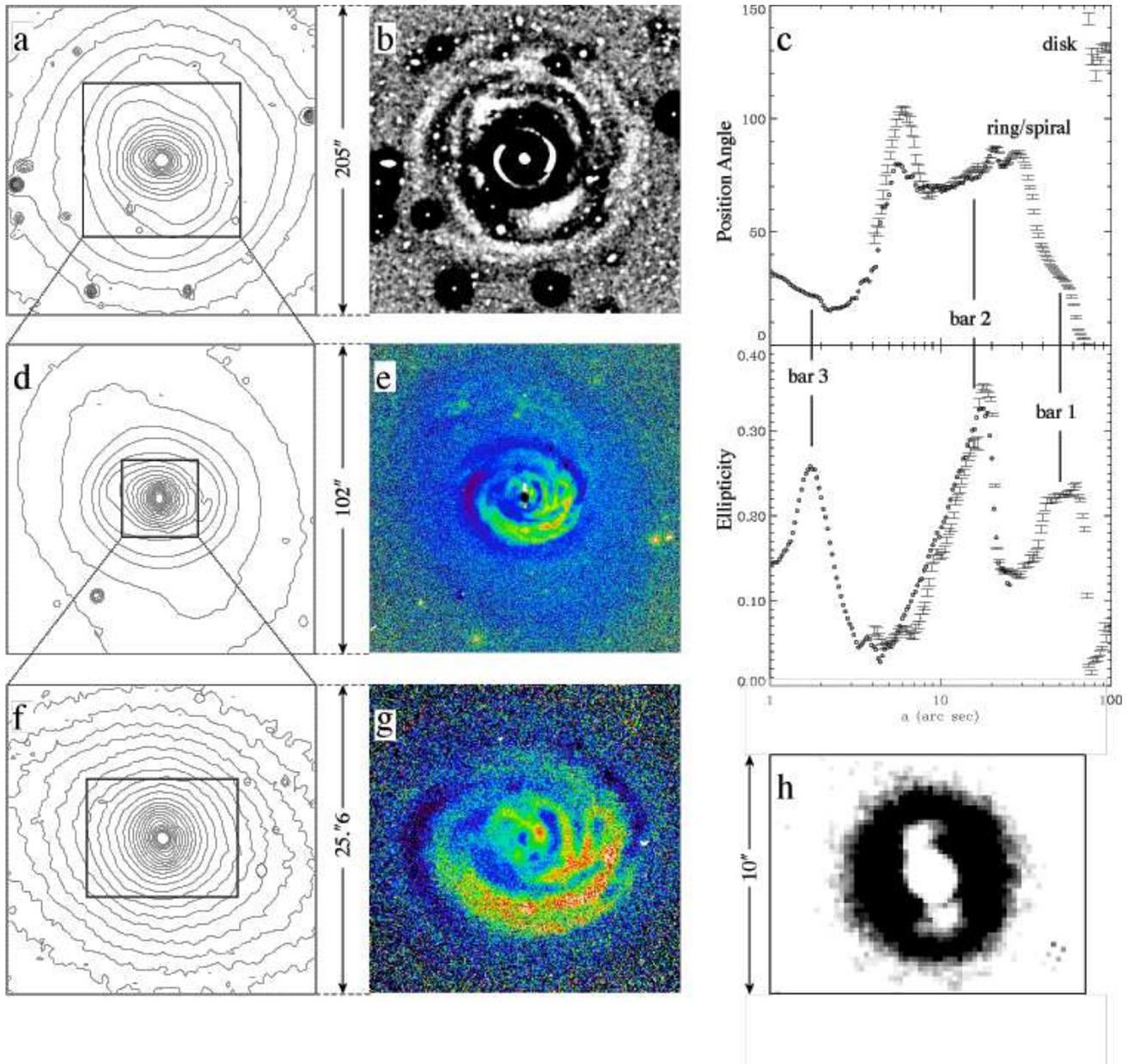}}

\caption{North is up and east is to the left in all images; no
deprojection has been attempted.  NGC 2681: \textbf{a} WIYN R-band
image: contour plot of the square root of logarithmic brightness,
showing outer disk, primary and secondary bars.  \textbf{b} Unsharp
mask, showing inner and outer rings; dark circles are due to
foreground stars.  \textbf{c} Isophotal ellipse fits to WIYN R-band
(error bars) and NICMOS3 F160W (circles) images.  \textbf{d} R-band
image, logarithmic brightness contours.  \textbf{e} Median-smoothed
\protect\br{} color map.
The nucleus is affected by saturation.  \textbf{f} NICMOS3 F160W
image, showing tertiary bar.  \textbf{g} WIYN \protect\br{} colormap
(no saturation, unsmoothed) for the same region; the blue center is
probably the LINER nucleus. \textbf{h} Unsharp mask of boxed region of
panel f, showing possible spiral arms outside the tertiary bar.}

\end{figure}

\clearpage

\begin{figure}
\scalebox{0.648}{\includegraphics{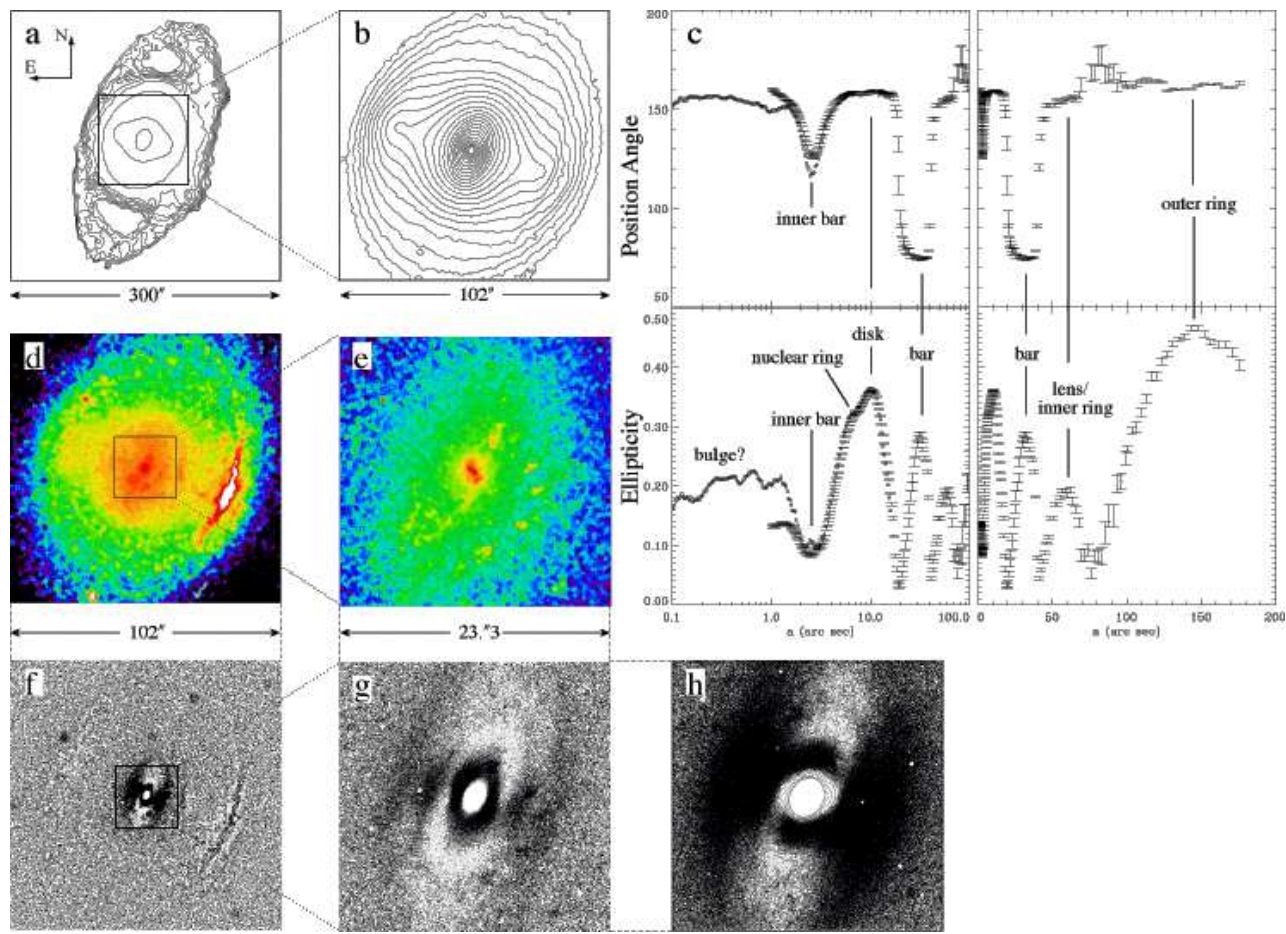}}
\caption{NGC 3945: \textbf{a} and \textbf{b} WIYN R-band images,
logarithmically scaled.  \textbf{c} Isophotal ellipse fits to WIYN
R-band (error bars) and WFPC2 F814W (circles) images: left, on a
logarithmic scale; right, on a linear scale. \textbf{d}
Median-smoothed WIYN \protect\br{} color map shows dust in the inner
ring and inner disk. \textbf{e} WFPC2 \protect\vi{}
(F555W$\!-\!$F814W) color map, showing dust clumps west of the nucleus
and inside the nuclear ring; the ring itself is not apparent.
\textbf{f} Unsharp mask of WIYN B-band image on the same scale as b,
showing the dusty inner ring on both sides of the main bar.
\textbf{g} Unsharp mask of F814W image, showing the nuclear ring and
short dust lanes in the western disk.  \textbf{h} as for g but with a
higher level of smoothing used for the mask, to show larger scale
features better.  Contours show the partly spiral nature of the inner
ring, crossed by an apparent inner bar.}
\end{figure}

\clearpage

\begin{figure}
\scalebox{0.648}{\includegraphics{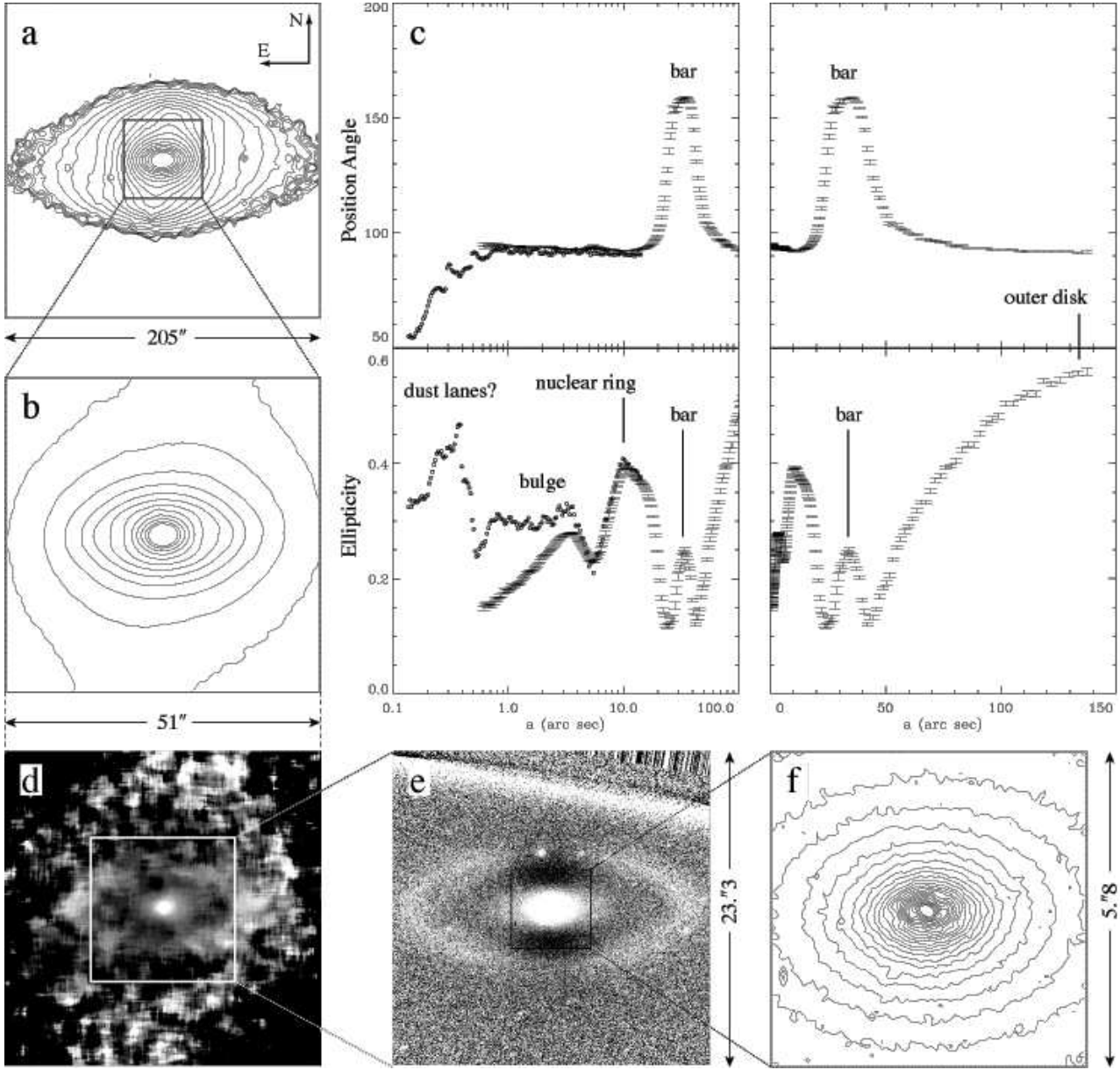}}
\caption{NGC 4371: \textbf{a} WIYN R-band image: contour levels are
square root of logarithmic brightness.  \textbf{b} WIYN R-band image,
logarithmic contours; note boxy isophotes inside the ring ($a <
10\protect\arcsec$).  \textbf{c} Ellipse fits to isophotes for WIYN
R-band (error bars) and WFPC2 F555W (circles) images.  \textbf{d}
Median-smoothed \protect\br{} color map from WIYN images; lighter
areas are redder.  The
apparent ring at $a \approx 15\protect\arcsec$ is at most 0.05
magnitudes redder in \protect\br{} than the nuclear ring. \textbf{e}
Unsharp mask of the F555W image, showing the nuclear ring, with
apparent ellipticity $\approx 0.6$ and $a \approx 10\protect\arcsec$.
The bright line crossing at the top is the edge of the PC2 chip.
\textbf{f} Twisted isophotes in the center of the WFPC2 image, perhaps
due to dust.}
\end{figure}

\clearpage

\end{document}